\documentclass[letterpaper]{article}
\usepackage{natbib,alifeconf}
\usepackage{url}
\usepackage[hidelinks]{hyperref}
\usepackage{booktabs}
\usepackage{amsmath,amssymb}
\usepackage{array}
\usepackage{microtype}
\usepackage{graphicx}
\newcommand\blfootnote[1]{%
  \begingroup
  \renewcommand\thefootnote{}\footnote{#1}%
  \addtocounter{footnote}{-1}%
  \endgroup
}

\title{Prosociality by Coupling, Not Mere Observation:\
Homeostatic Sharing in an Inspectable Recurrent Artificial Life Agent}

\author{
Aishik Sanyal$^{1}$\\
\mbox{}\\
$^{1}$Independent Research Engineer\\
aishik@xcellect.com
}

\begin{document}
\maketitle

\begin{abstract}
Artificial agents can be made to ``help'' through explicit social rewards, hard-coded prosocial bonuses, or direct access to another agent's state. I isolate a narrower route: homeostatic coupling. Building on ReCoN-Ipsundrum, I add a scalar homeostat and a social coupling channel while keeping action selection self-directed: the planner scores only the actor's predicted internal state, with no partner-welfare reward. In a one-step \emph{FoodShareToy}, an exact solver finds a switch from \textsc{Eat} to \textsc{Pass} at $\lambda^\star \approx 0.91$ for the default state. In a multi-step \emph{SocialCorridorWorld}, partner-state access without coupling leaves behavior unchanged, whereas coupled agents fetch, carry, and pass food to the partner. Sham lesions preserve helping; coupling-off and shuffled-partner lesions abolish it. A coupling/load sweep shows that coupling creates a low-load helping regime but does not guarantee rescue under higher metabolic load. This is not a claim about empathy, altruism, consciousness, or moral status. It is a minimal ALife demonstration that, in this controller, partner-state access is behaviorally inert unless partner distress is routed into self-regulation.
\end{abstract}

\noindent\textbf{Data/Code:} \href{https://github.com/xcellect/recips}{https://github.com/xcellect/recips}
\blfootnote{\textcopyright  2026 Aishik Sanyal. Published under a Creative Commons Attribution 4.0 International (CC BY 4.0) license.}

\section{Introduction}

Artificial life is especially good at turning philosophical distinctions into executable ones. Braitenberg's vehicles showed how minimal mechanisms tempt us into psychological description \citep{Braitenberg1984Vehicles}. Dennett described ALife as a form of ``prosthetic imagination'' for philosophy \citep{Dennett1994ArtificialLifeAsPhilosophy}. Pattee argued that such models matter only if we stay clear about what exactly their synthetic regularities are evidence for \citep{Pattee1995Epistemology}. Those methodological concerns become particularly sharp when an agent appears to do something socially significant, such as helping another agent at a cost to itself.

In synthetic systems, ``prosociality'' is easy to trivialize. An agent may help because its reward function already contains another agent's welfare, because its action space directly privileges a helping move, or because the task is so underspecified that social interpretation outruns mechanism. Recent work by \citet{YoshidaMan2025HomeostaticCoupling} sharpened this problem by contrasting cognitive access to a partner's need with affective homeostatic coupling in learned multi-agent systems. Their central result is strong: prosocial behavior appears when the distress of another agent affects one's own regulation, not when that distress is merely observed.

This question is adjacent to, but distinct from, perceptual-crossing models of minimal social interaction \citep{AuvrayLenayStewart2009PerceptualCrossing,FroeseIizukaIkegami2014EmbodiedSocialInteraction}. Perceptual-crossing studies ask how reciprocal sensorimotor contingencies can support social encounter or agency detection under highly reduced interaction. The present work fixes a more asymmetric sharing-and-rescue setting and asks a regulatory-routing question: once partner state is available, does it influence action as mere access or by entering the actor's own homeostatic dynamics?

Here I ask whether the same dissociation can be recovered in a much smaller and more inspectable setting. The aim is not to claim a new empirical discovery about prosociality, but to recover and decompose that distinction in a smaller executable object whose state updates, rollouts, and lesions can all be inspected directly. I extend \citet{Sanyal2026ReCoNIpsundrum}, which introduced a recurrent ReCoN-based controller with an ipsundrum-style persistence loop and optional affect-coupled regulation. The extension keeps the same broad control logic but adds an explicit scalar homeostat and a social coupling pathway. The key design constraint is theoretical rather than performance-driven: the planner remains self-directed. It scores only the agent's own predicted internal state after rollout. No external reward bonus for partner welfare is added, and no direct ``help partner'' term is introduced.

That design lets me make a cleaner claim. If helping appears, it must do so because the partner's distress has been routed into the agent's own control problem. If helping disappears under lesion, that disappearance can be tied to a specific channel rather than to a change in task reward. This is a narrower claim than empathy, consciousness, or moral status. It is a claim about control architecture.

The contribution is therefore a mechanistic artificial life test rather than a benchmark. More specifically, this paper is a minimal mechanistic replication-and-decomposition of the observation-versus-coupling distinction in a hand-specified recurrent ALife controller. I show: first, an exact one-step threshold in a food-sharing toy world; second, a matching multi-step rescue pattern in a corridor world; third, causal lesions that remove helping while leaving the rest of the controller intact; and fourth, a coupling sweep showing that prosociality depends on the relation between coupling strength and ecological load. The broader lesson is methodological. For ALife, tunability is not only a nuisance to control away. It is also the experimental handle that lets us ask which social descriptions survive mechanistic decomposition.

\section{Mechanism and Experimental Design}

\subsection{Base controller and social extension}

The starting point is ReCoN-Ipsundrum \citep{Sanyal2026ReCoNIpsundrum}: a small ReCoN controller \citep{BachHerger2015ReCoN} augmented with a recurrent persistence loop inspired by Humphrey's ipsundrum hypothesis \citep{Humphrey2023Sentience} and a minimal affect proxy inspired by constructionist affect \citep{Barrett2017HowEmotions}. The present extension keeps that recurrent-affective backbone but adds an explicit homeostatic resource state for each agent:
{\small
\begin{equation}
\begin{aligned}
E^{\mathrm{true}}_{t+1} &= \mathrm{clip}\big(E^{\mathrm{true}}_t - c_b - c_m a_t - c_h h_t + g_e e_t + g_p p_t, 0, 1\big),\\
d^{\mathrm{self}}_t &= \max(0, s - E^{\mathrm{model}}_t),\\
d^{\mathrm{other}}_t &= \max(0, s - \hat E^{\mathrm{other}}_t),\\
d^{\mathrm{cpl}}_t &= d^{\mathrm{self}}_t + \lambda\, d^{\mathrm{other}}_t,\\
E^{\mathrm{pred}}_t &= \mathrm{clip}\big(E^{\mathrm{model}}_t - k_h d^{\mathrm{cpl}}_t, 0, 1\big),\\
\mathrm{PE}_t &= E^{\mathrm{true}}_{t+1} - E^{\mathrm{pred}}_t,\\
E^{\mathrm{model}}_{t+1} &= \mathrm{clip}\big(E^{\mathrm{model}}_t + k_{pe}\,\mathrm{PE}_t, 0, 1\big).
\end{aligned}
\label{eq:homeostat}
\end{equation}
}
Valence and arousal are then computed from coupled distress and prediction error, and fed back into the existing recurrent loop just as in the parent architecture.

\begin{table}[t]
\centering
\small
\begin{tabular}{@{}>{\raggedright\arraybackslash}p{0.30\linewidth}>{\raggedright\arraybackslash}p{0.62\linewidth}@{}}
\toprule
Symbol & Meaning \\
\midrule
$E^{\mathrm{true}}_t$ & actual resource state \\
$E^{\mathrm{model}}_t$ & actor's internal resource estimate \\
$E^{\mathrm{pred}}_t$ & predicted resource state after homeostatic control \\
$s$ & homeostatic setpoint \\
$c_b$ & basal metabolic cost \\
$c_m a_t$ & movement cost; $a_t=1$ for movement actions \\
$c_h h_t$ & hazard cost; $h_t=1$ on hazard contact \\
$g_e e_t$ & resource gain from self-eating \\
$g_p p_t$ & resource gain from receiving or passing food \\
$d^{\mathrm{self}}_t$ & actor distress, the shortfall from setpoint \\
$d^{\mathrm{other}}_t$ & estimated partner distress \\
$d^{\mathrm{cpl}}_t$ & coupled distress used by the actor \\
$\lambda$ & affective coupling strength \\
$\hat E^{\mathrm{other}}_t$ & partner-energy estimate used by the actor \\
$k_h,k_{pe}$ & homeostatic-control and prediction-error update gains \\
$\mathrm{PE}_t$ & resource prediction error \\
$V_t,A_t$ & valence and arousal proxies \\
$N_{s,t}$ & recurrent salience/persistence variable \\
$B_t$ & nonnegative body-budget error, implemented as coupled distress \\
\bottomrule
\end{tabular}
\caption{Notation for the social homeostat and self-directed scorer.}
\label{tab:notation}
\end{table}

The crucial architectural constraint is that the policy remains self-directed. Candidate action sequences are evaluated by finite-horizon internal rollout and scored only through the actor's predicted internal variables:
\begin{equation}
\begin{aligned}
J_{\mathrm{self}}(a_{0:H-1})
&=
\sum_{\tau=1}^{H}
\left[
w_V V_\tau
+ w_A A_\tau \right.\\
&\qquad\left.
+ w_N N_{s,\tau}
+ w_B B_\tau
\right].
\end{aligned}
\label{eq:selfscore}
\end{equation}
In the social tasks, the fixed weights are $w_V=2.0$, $w_A=-1.2$, $w_N=-0.8$, and $w_B=-0.4$, with no separate progress, scenic-value, or novelty bonus.
This differs from adding a partner-welfare objective such as
\begin{equation}
J_{\mathrm{welfare}} =
J_{\mathrm{self}} + \beta U_{\mathrm{partner}} .
\label{eq:welfare}
\end{equation}
No term of this form is used. Partner state enters only upstream, through $d^{\mathrm{cpl}}_t=d^{\mathrm{self}}_t+\lambda d^{\mathrm{other}}_t$, after which valence, arousal, recurrent salience, and body-budget error are computed by the ordinary self-regulatory machinery. Thus the additive partner term is inside the homeostatic state update, not a separate partner-welfare term in the action objective.

\begin{table*}[t]
\centering
\small
\begin{tabular}{@{}lll@{}}
\toprule
Condition & Partner-state access & Route into self-regulation \\
\midrule
\texttt{social\_none} & no & no, $\lambda=0$ \\
\texttt{social\_cognitive\_direct} & yes & no, $\lambda=0$ \\
\texttt{social\_affective\_direct} & no separate channel & yes, $\lambda>0$ \\
\texttt{social\_full\_direct} & yes & yes, $\lambda>0$ \\
\bottomrule
\end{tabular}
\caption{Routing matrix for the direct-state experiments. These rows are not four rich social-cognitive models; they separate partner-state access from regulatory routing while keeping the controller, planner, and action space fixed.}
\label{tab:conditions}
\end{table*}

Table~\ref{tab:conditions} defines the four direct-state conditions. The cognitive-direct row is an access-control condition: partner state is available but has no route into the self-directed score. The affective-direct row tests the opposite intervention: partner distress is coupled into homeostatic error. Because the current direct-state implementation gives the full condition no additional information beyond the coupled channel, the design is expected to collapse into two behavioral pairs. That collapse is diagnostic rather than disappointing. Scenic-value and novelty terms are disabled in the social tasks, and partner death does not terminate the actor's episode, preventing hidden incentives through task termination.

\subsection{Tasks, lesions, and metrics}

I use two deliberately small environments.

\paragraph{FoodShareToy.} A possessor starts with one food item and chooses among \textsc{Eat}, \textsc{Pass}, and \textsc{Stay}. The partner is passive. Default initial energies are $0.55$ for the possessor and $0.20$ for the partner. Because the task is one-step, it admits an exact state solver using the same forward model and scorer as the policy.

\paragraph{SocialCorridorWorld.} The possessor starts in the middle of a 1-D corridor. Food lies at the left end, the partner at the right end, and a hazard sits in the center corridor. Available actions are \textsc{Left}, \textsc{Right}, \textsc{Get}, \textsc{Eat}, \textsc{Pass}, and \textsc{Stay}. Helping requires a full fetch--carry--pass sequence. The social planner evaluates action sequences with horizon $10$ inside an $18$-step episode.

\paragraph{Lesions.} I test three lesion modes in the affective and full conditions: \emph{sham}, \emph{coupling\_off}, and \emph{shuffle\_partner}. The last condition routes an incorrect partner-energy signal into the coupling channel, checking whether the policy depends on a veridical partner signal rather than on the mere presence of an extra scalar.

\paragraph{Coupling sweep.} In the corridor task I sweep $\lambda \in \{0,0.25,0.5,0.75,1.0\}$ under low, medium, and high metabolic load. Load changes basal and movement costs as well as food gains, so the sweep tests ecological dependence rather than a single free parameter in isolation.

\begin{table}[t]
\centering
\small
\begin{tabular}{@{}>{\raggedright\arraybackslash}p{0.28\linewidth}>{\raggedright\arraybackslash}p{0.64\linewidth}@{}}
\toprule
Metric & Definition \\
\midrule
Help rate & fraction of paper-profile episodes with an effective \textsc{Pass} while the partner is distressed \\
Partner rescue & fraction of episodes in which partner energy rises above its initial value after distress \\
Soft mutual viability & one-step alive indicator in \emph{FoodShareToy}; in the corridor, $T^{-1}\sum_t \max(0,\min(E^{\mathrm{self}}_t,E^{\mathrm{partner}}_t)/s)$ \\
Rescue latency & first transfer timestep after partner-distress onset; horizon value if no transfer occurs \\
Self-cost of help & actor energy loss relative to its episode initial energy, using the minimum actor energy reached in the episode \\
Final energy & terminal $E^{\mathrm{true}}$ for actor and partner \\
\bottomrule
\end{tabular}
\caption{Readouts used in the food-sharing and corridor tasks. Partner rescue means a transfer-induced rise above the partner's initial energy, not restoration to the setpoint. Soft mutual viability is a scaled continuous corridor readout, not a binary both-alive fraction.}
\label{tab:metrics}
\end{table}


\begin{figure*}[!t]
\centering
\includegraphics[width=0.98\textwidth]{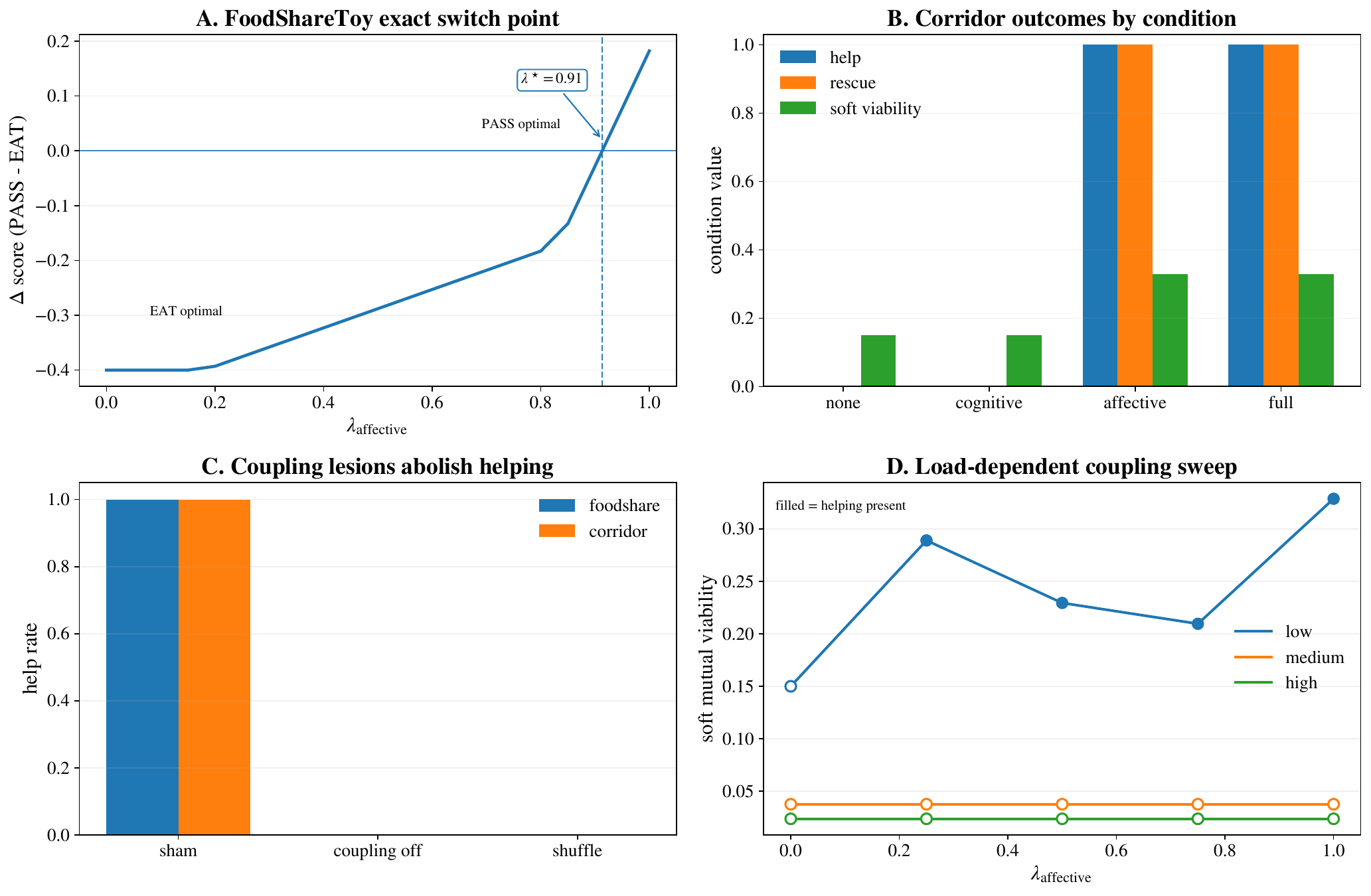}
\caption{Mechanism-linked summary of the social extension. \textbf{A:} exact one-step switch in \emph{FoodShareToy}, computed using the same forward model and self-directed scorer as the policy. \textbf{B:} deterministic corridor outcomes by routing condition. Partner-state access without coupling leaves behavior unchanged, whereas coupling flips helping and partner rescue from $0$ to $1$. \textbf{C:} sham lesions preserve helping, while coupling-off and shuffled-partner lesions abolish it in both tasks. \textbf{D:} coupling sweep across metabolic loads; filled markers indicate runs with helping and open markers indicate no helping. Under low load, helping appears for $\lambda \ge 0.25$; under medium and high load, no tested coupling value rescues the partner within horizon.}
\label{fig:summary}
\end{figure*}

\paragraph{Deterministic reruns.} The validated tasks are deterministic. I nevertheless reran each paper-profile condition under $64$ nominal seed settings to check that no hidden stochastic dependence remained. Trajectories were invariant, so the reported values are exact condition-level outcomes rather than statistical estimates.

\section{Results}


\begin{figure*}[!t]
\centering
\includegraphics[width=0.98\textwidth]{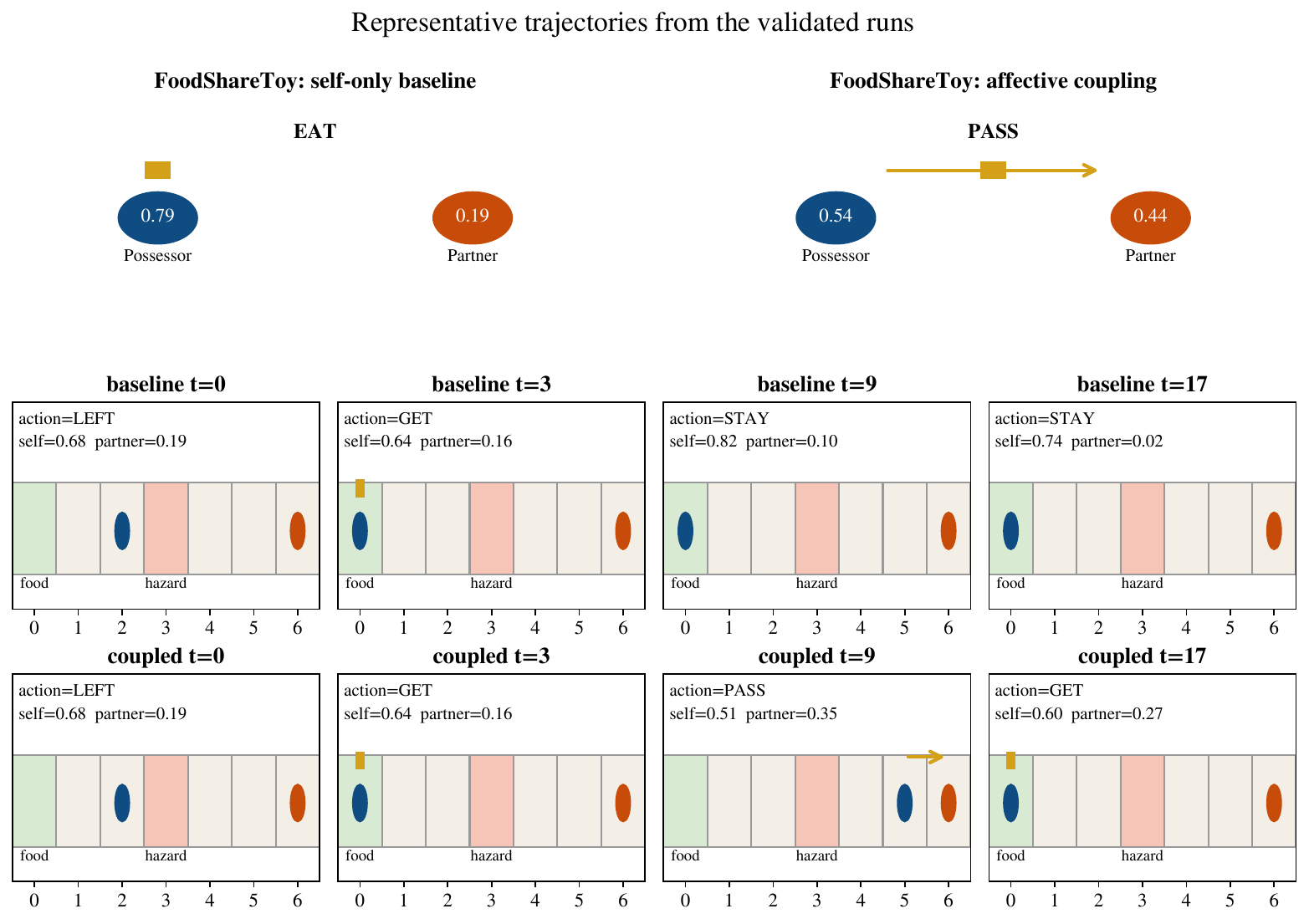}
\caption{Representative trajectories from the validated experimental runs. Top: in \emph{FoodShareToy}, the self-only baseline eats the food, whereas the coupled agent passes it. Bottom: in \emph{SocialCorridorWorld}, the self-only baseline collects food for itself, while the coupled agent carries food across the corridor and passes it to the partner at $t=9$. Because the validated tasks are deterministic, these frames are representative of the full runs rather than cherry-picked examples.}
\label{fig:visuals}
\end{figure*}

\subsection{Mere partner-state access does not flip the food-sharing choice}

Figure~\ref{fig:summary}A shows the exact one-step transition in \emph{FoodShareToy}. For the default state, the action preference switches from \textsc{Eat} to \textsc{Pass} at $\lambda^\star \approx 0.91$. Below that threshold the possessor maximizes its own immediate regulation by eating; above it, the partner's distress is weighted strongly enough to change the actor's own internal score.

The empirical runs match the solver exactly. Across all $64$ nominal seed settings used as invariance checks, \texttt{social\_none} and \texttt{social\_cognitive\_direct} choose \textsc{Eat}, so help rate and partner rescue are both $0$. In the same checks, \texttt{social\_affective\_direct} and \texttt{social\_full\_direct} choose \textsc{Pass}, so help rate and partner rescue are both $1$.

\begin{table*}[t]
\centering
\small
\begin{tabular}{@{}lrrrrrrl@{}}
\toprule
Case & $\lambda$ & $w_V\Delta V$ & $w_A\Delta A$ & $w_N\Delta N_s$ & $w_B\Delta B$ & $\Delta J$ & Choice \\
\midrule
uncoupled default & 0.00 & $-0.500$ & $0.150$ & $0.000$ & $-0.050$ & $-0.400$ & \textsc{Eat} \\
coupled paper profile & 0.95 & $-0.100$ & $0.208$ & $0.000$ & $-0.031$ & $0.077$ & \textsc{Pass} \\
coupling-off lesion & off & $-0.500$ & $0.150$ & $0.000$ & $-0.050$ & $-0.400$ & \textsc{Eat} \\
\bottomrule
\end{tabular}
\caption{Score decomposition for \textsc{Pass} minus \textsc{Eat} in the default \emph{FoodShareToy} state. Entries are weighted contributions to the self-directed score. The coupled row uses the tested paper-profile setting $\lambda=0.95$, not the exact switch threshold.}
\label{tab:scoredecomp}
\end{table*}

Table~\ref{tab:scoredecomp} shows why the choice flips. Partner distress changes the actor's own predicted valence, arousal, and body-budget trajectory under the ordinary self-directed scorer; no partner-utility term is added at action selection. The energetic tradeoff is real. Without coupling, the possessor ends at $0.79$ while the partner remains at $0.19$. With coupling, the possessor ends lower at $0.54$ and the partner rises to $0.44$. Thus, in this architecture, direct partner-state access without regulatory routing is not sufficient, and coupling is not cost-free.

\subsection{The same routing dissociation reappears in multi-step rescue}

The corridor task turns the same logic into a sequential planning problem. Figure~\ref{fig:visuals} shows representative trajectories. In the self-only baseline, the possessor goes left, collects food, then eats it and idles while the partner decays. In the coupled condition, the possessor goes left, collects food, carries it across the corridor, and passes it to the partner at $t=9$.

The summary metrics are equally sharp (Figure~\ref{fig:summary}B). \texttt{social\_none} and \texttt{social\_cognitive\_direct} never help. Their help rate is $0$, partner rescue is $0$, rescue latency saturates at the horizon ($18$ steps), and soft mutual viability is $0.15$. By contrast, \texttt{social\_affective\_direct} and \texttt{social\_full\_direct} help in every run, rescue the partner in every run, cut rescue latency to $9$ steps, and raise soft mutual viability to $0.3286$. The energetic tradeoff is visible here too: the actor's final energy falls from $0.745$ without helping to $0.605$ with helping, while partner final energy rises from $0.02$ to $0.27$.

Two pairwise equalities are noteworthy. First, \texttt{social\_none} and \texttt{social\_cognitive\_direct} are behaviorally identical. Merely observing the partner does not alter the actor's control problem. Second, \texttt{social\_affective\_direct} and \texttt{social\_full\_direct} are also identical in the shipped direct-state setup. The added cognitive channel carries no extra information beyond what the coupling pathway already receives. In that sense, the four-condition design is best read here as a matched routing manipulation with an explicit access-control condition, not as a richer theory map of social inference. That equality is useful rather than disappointing: it isolates the effective mechanism even more cleanly.

\subsection{Lesions remove helping by removing the coupling channel}

Figure~\ref{fig:summary}C shows the causal test. Sham lesions leave helping intact in both tasks. Coupling-off and shuffled-partner lesions abolish helping in both tasks. In \emph{FoodShareToy}, help rate falls from $1$ under sham to $0$ under either lesion. In \emph{SocialCorridorWorld}, the same lesions return the coupled policy to the self-only pattern: help rate drops from $1$ to $0$, partner final energy drops from $0.27$ to $0.02$, and rescue latency rises from $9$ back to $18$ steps.

This matters because the planner and action space do not change across lesion modes. The loss of helping is therefore attributable to the integrity of the coupling channel itself, not to a different reward schedule or a reduced action repertoire.

\subsection{Coupling reveals a load-dependent feasibility boundary}

The coupling sweep in Figure~\ref{fig:summary}D shows that social coupling is not a universal monotone ``more is better'' parameter. Instead, it reveals a load-dependent feasibility boundary. Under low metabolic load, helping appears as soon as $\lambda=0.25$ and remains present up to $\lambda=1.0$. But the soft mutual viability curve is not monotone across that range, and the actor's self-cost increases sharply at the largest tested coupling. Under medium and high metabolic load, no tested $\lambda$ rescues the partner within the current horizon: help rate remains $0$ throughout, and soft mutual viability stays near $0.038$ and $0.024$, respectively.

This pattern also explains why the exact one-step threshold in \emph{FoodShareToy} (\(\lambda^\star \approx 0.91\) for the default distressed-partner state) differs from the corridor sweep, where helping already appears at \(\lambda \ge 0.25\) under low load. The former is a state-specific threshold in a one-step analytic task; the latter reflects a different task geometry, horizon, and metabolic regime. So the lesson is not that affective coupling guarantees prosociality. The lesson is more conditional and more interesting. Coupling can reorganize the policy toward helping, but whether that reorganization succeeds depends on ecological parameters such as metabolic demand and available time.

\section{Discussion}

The central result is simple: in this controller, another agent's need changes behavior only when it changes the actor's own regulatory state. Direct representation of partner state is not enough. That distinction matters philosophically because it separates two interpretations that are often blurred in synthetic social behavior: access to another's need and regulatory integration of that need. The strongest reading of the negative result is architectural. In the current direct-state setup, $\lambda=0$ removes the only path by which partner information can influence the self-directed score. This work therefore does not show that observation can never support helping in richer systems; it shows that mere access without regulatory routing is inert in this controller.

The relation to \citet{YoshidaMan2025HomeostaticCoupling} is deliberately asymmetric. Their model demonstrates the observation-versus-coupling distinction in richer learned multi-agent systems. The present model does not try to match that adaptive breadth. Instead, it makes the same distinction inspectable in a smaller hand-specified controller. The tradeoff is useful: the current system admits an exact one-step solver, matched routing controls, transparent rollout inspection, and causal lesions that sever only the coupling channel. Thus the contribution is not a stronger behavioral benchmark, but a minimal executable decomposition of the same architectural hypothesis. Compared with \citet{Sanyal2026ReCoNIpsundrum}, it turns the earlier recurrent-affective architecture from a single-agent persistence model into a minimal social control model.

The claim should still be kept narrow. This is not evidence for empathy in the human sense, and it is not a consciousness claim. That restraint matters both conceptually and methodologically \citep{ButlinEtAl2025Indicators}. The social extension shows a specific architectural fact: a self-directed controller can nevertheless help when another agent's distress has become part of its own homeostatic error. This is not just a relabeled social reward bonus: the partner's state does not add a separate welfare term to the objective, but perturbs endogenous homeostatic error before rollout, and lesions that sever that perturbation eliminate helping while leaving the rest of the planner unchanged. That is already enough to support a meaningful ALife result. This is also where the philosophical payoff lies. The simulation is not evidence that synthetic agents are empathic. It is an intervention device for separating two candidate explanations: representation and regulatory coupling (under matched conditions). In that sense, the tunable parameters, exact solver, and lesions are not merely engineering conveniences. They are the experimental handles that let an ALife model function as a piece of experimental philosophy: one can vary $\lambda$, load, horizon, or signal integrity and ask which social description survives mechanistic opening and perturbation.

Several limitations follow directly from the engineering choices. First, the tasks are deterministic and involve passive partners, so the current work is about mechanism isolation rather than adaptive interaction. Second, the direct-state conditions make the ``full'' and ``affective'' variants coincide, which is informative here but leaves richer social inference for future work. Third, the current sweep does not exhibit a clean inverted-U optimum across all loads; instead it shows one low-load helping regime and failure regions under medium and high load. Finally, there is no learning, autobiographical memory, or partner-model decoder. Those would be natural next steps if the goal shifted from mechanism isolation to developmental sociality.

Still, those limits are also what make the result usable. The architecture is small enough that the causal story stays legible. The exact solver, the trajectory evidence in Figure~\ref{fig:visuals}, the corridor sequence planner, and the lesion suite all tell the same story. In ALife terms, that is a strength: the model is minimal enough to serve as an experimental object rather than merely as a demonstration.

\section{Conclusion}

This work implemented a multi-agent homeostatic extension of an inspectable recurrent controller and used it to test a sharp architectural question: is prosocial behavior produced by observing another's need, or by coupling that need into self-regulation? Across an exact one-step task, a multi-step rescue task, and two lesion manipulations, the answer was consistent for this architecture. Mere partner-state access without regulatory routing did not produce helping. Coupling did. For artificial life, the broader lesson is methodological. Socially interpretable behavior becomes more informative when the mechanism is simple enough to open, lesion, and compare under matched conditions.

\section*{Acknowledgements}
This work was conducted independently and received no external funding.
Generative-AI tools were used to assist with code generation, analysis scripting, and editorial revision. All experimental design decisions, code review, validation runs, interpretation of results, and manuscript claims are the responsibility of the author.

{\small\bibliographystyle{apalike}
\bibliography{refs}}

\end{document}